\documentclass[12pt]{article}
\usepackage{amsmath}
\usepackage{graphicx,psfrag,epsf}
\usepackage{enumerate}
\usepackage{natbib}
\usepackage{url} 

\newcommand{\blind}{0}

\begin{document}

\def\spacingset#1{\renewcommand{\baselinestretch}%
{#1}\small\normalsize} \spacingset{1}


\if0\blind
{
  \title{\bf Estimating the treatment effect in a subgroup defined by an
  early post-baseline biomarker measurement in randomized
  clinical trials with time-to-event endpoint.}
  \author{Bj\"{o}rn Bornkamp\hspace{.2cm}\\
    and \\
    Georgina Bermann \\
    Clincal Development and Analytics, Novartis Pharma AG}
  \maketitle
} \fi

\if1\blind
{
  \bigskip
  \bigskip
  \bigskip
  \begin{center}
    {\LARGE\bf Estimating the treatment effect in a subgroup defined by an
  early post-baseline biomarker measurement in randomized
  clinical trials with time-to-event endpoint.}
\end{center}
  \medskip
} \fi

\bigskip
\begin{abstract}
  Biomarker measurements can be relatively easy and quick to obtain
  and they are useful to investigate whether a compound works as
  intended on a mechanistic, pharmacological level.  In some
  situations, it is realistic to assume that patients, whose
  post-baseline biomarker levels indicate that they do not
  sufficiently respond to the drug, are also unlikely to respond on
  clinically relevant long term outcomes (such as time-to-event).
  However the determination of the treatment effect in the subgroup of
  patients that sufficiently respond to the drug according to their
  biomarker levels is not straightforward: It is unclear which
  patients on placebo would have responded had they been given the
  treatment, so that naive comparisons between treatment and placebo
  will not estimate the treatment effect of interest.  The purpose of
  this paper is to investigate assumptions necessary to obtain causal
  conclusions in such a setting, utilizing the formalism of causal
  inference. Three approaches for estimation of subgroup effects will
  be developed and illustrated using simulations and a case-study.
\end{abstract}

\noindent%
\vfill

\newpage
\spacingset{1.45} 

\section{Introduction}
\label{sec:intro}

Biomarker measurements can be relatively easy and quick to obtain, and
are useful to investigate whether a compound works as intended on a
mechanistic, pharmacological level. In some situations, it is
plausible to assume that post-baseline biomarker responders are also
more likely to respond better on clinically relevant long term
outcomes, which are often of time-to-event type.

This research is motivated by the CANTOS outcome study in prevention
of cardiovascular events \citep{ridker:2017}. Inflammation has been
identified as playing a key role in atherosclerosis, for example in
the formation and rupture of atherosclerotic plaques
\citep{hans:2005}.  The CANTOS trial investigated canakinumab, an
anti-inflammatory agent, against placebo. The primary outcome was the
time to a major adverse cardiac (MACE) event, a composite endpoint
consisting of cardiovascular death, non-fatal myocardial infarction
and stroke, and was statistically significant. In this specific case
the biomarker of interest is a downstream inflammatory marker high
sensitivity C-reactive protein (hs-CRP), where lower values indicate
less inflammation. Interest focuses on determination of the treatment
effect for hs-CRP patients that, 3 months after treatment start, were
able to lower hs-CRP below a specific target level.

The determination of the treatment effect in a subgroup of patients
that is defined based on post-baseline biomarker levels in the
treatment group (e.g. indicated by reaching biomarker levels smaller
than some threshold) is not straightforward: The biomarker might have
a prognostic effect on the outcome, independent of treatment. For
example hs-CRP is a known prognostic risk factor for cardiovascular
events. It is likely that patients who reach the biomarker target
level on treatment also had a better (i.e. smaller) biomarker
measurement at baseline compared to patients, who do not reach the
target. A naive comparison of the biomarker responder subgroup on
treatment, to the complete placebo group will thus likely overestimate
the treatment effect. Similarly a naive comparison of the biomarker
responder subgroup patients on treatment, against the biomarker
responder subgroup patients on placebo will also likely be
biased. Patients, who reach biomarker levels below the target on
placebo are likely more ``healthy'' than the biomarker responders on
treatment. So such a comparison would likely underestimate the
treatment effect.

The purpose of this paper is to investigate the assumptions necessary
to draw causal conclusions, using the formal language of causal
inference developed over the past decades, see for example
\cite{pear:2009}, \cite{imbe:rubi:2015} or \cite{hern:robi:2018} for
reviews. ``Valid conclusions'' here means that the treatment effect
should be attributable to the difference in treatment and not to
differences in the two compared populations.

In the causal inference literature there are several papers dealing
with related questions in more detail. The paper \cite{fran:rubi:2002}
introduced the term principal stratification. A reference providing a
review of many of the recent approaches and developing new approaches
is \cite{ding:lu:2017}. Related references among others are for
example \cite{joff:smal:hsu:2007, schw:li:meal:2011, zigl:beli:2012,
  jo:stua:2009,jo:stua:2011,stua:jo:2015,kern:stua:hill:2015}.

The purpose of this article is to provide a review of the underlying
problem and propose methods that could be utilized in this setting. In
Section \ref{sec:methods} we utilize causal inference techniques to
express the estimands of interest in terms of quantities that are
identifiable in a randomized clinical trial. In Section \ref{sec:est}
we discuss how these quantities can be estimated. In Section
\ref{sec:sims} we perform a simulation study to illustrate the
proposed methods in a specific situation. Section \ref{sec:appl}
illustrates the method for a specific simulated data set. Section
\ref{sec:concl} concludes.

\section{Methods}
\label{sec:methods}

\subsection{Estimand of interest}
\label{sec:estimand}

Assume a randomized trial with a treatment and a placebo arm. Let $X$
denote the treatment indicator, denoting whether a patient was
randomized to treatment ($x=1$) or placebo ($x=0$). Further let
$\beta(x)$ denote the potential continuous biomarker outcome under
treatment $x$ at a specified time $\mathcal{T}$ after start of the
study. Further let $B(x)$ be the binary indicator, defined as whether
$\beta(x)<\beta^*$ for a target threshold $\beta^*$. Let $T(x)$ be the
potential event time under treatment $x$. Here we focus on composite
endpoints (a mixture of non-fatal events and death).  The
corresponding observed values will be denoted by $\beta = \beta(X)$,
$B = B(X)$ and $T = T(X)$, where $X$ is the treatment actually
obtained. Let $Z$ denote a vector of baseline variables influencing
the biomarker and the event time.  We assume in what follows that
$\mathcal{T}$ is small, so that no events are observed before
$\mathcal{T}$. We include a discussion in Appendix \ref{sec:ev} on how
to handle events before time $\mathcal{T}$.

As discussed in the introduction, the population of interest is the
subgroup of patients, that, if given the treatment would be biomarker
responders. In terms of the introduced notation these are the patients
with $B(1)=1$.

Different summary measures exist to define a treatment effect in terms
of time-to-event outcomes. We focus here on the difference in survival
probabilities at a given time-point $t$
\begin{equation}
  \label{eq:estimand1}
  \Delta(t)=P(T(1)>t|B(1)=1)-P(T(0)>t|B(1)=1),
\end{equation}
and the difference in restricted mean survival times
\citep{roys:parm:2011},
\begin{equation}
  \label{eq:estimand2}
  \Delta_{RMST,t^*}=\int_0^{t^*} \Delta(t) dt,
\end{equation}
integrated up to a time-point $t^*$. Here $P(T(1)>t|B(1)=1)$ and
$P(T(0)>t|B(1)=1)$ are the survival functions in the subgroup of
interest. 

Because biomarker response is a post-baseline event in a parallel
groups trial, we do not know which patients on placebo would be a
biomarker responder had they received treatment. Identification of the
estimand, and particular $P(T(0)>t|B(1)=1)$, hence requires
assumptions. In terms of the draft ICH E9 addendum \citep{ich:2017},
this is an example of a principal stratification estimand, where the
stratum of interest is the subgroup of the overall trial population
with $B(1)=1$.

\subsection{Identification of the Estimand}
\label{sec:notation}

As treatment $X$ is randomized we know that $X$ and the variables
$\{ T(x), \beta(x) , Z \},\; \; x=0,1$ are independent:
\begin{equation}
\label{eq:rand}
X \perp \{ T(x), \beta(x) , Z \}  \; \; x=0,1.
\end{equation}

In what follows we will concentrate on causal identification of
$P(T(1)>t|B(1)=1)$ and $P(T(0)>t|B(1)=1)$ as $\Delta(t)$ and
$\Delta_{RMST}(t^*)$ can be derived based on these.

Note that $P(T(1)>t|B(1)=1)$ can be identified from the observed trial
data because
\begin{eqnarray*}
\label{eq:act}
P(T(1)>t|B(1)=1)&=&P(T(1)>t|X=1, B(1)=1) \\
&=&P(T>t|X=1, B=1). 
\end{eqnarray*}
The first equality holds due to randomization (\ref{eq:rand}) and the
second equality holds due to the assumption of consistency, which
states that the distribution of $T(1)|X=1, B(1)=1$ is equal to the
distribution $T|X=1,B=1$ observed in the trial. The quantity
\begin{equation}
  \label{eq:pbob1}
 P(T(0)>t|B(1)=1),
\end{equation}
however cannot be estimated from the trial data without further causal
assumptions, because in a parallel groups trial, one cannot obtain the
survival time under placebo, for patients that were on treatment and
had $B=1$. In the following two sections, two different causal
assumptions will be explored that allow identification of this
quantity: One based on conditional independence assumptions using
baseline covariates $Z$ and the other is an analysis based on the
monotonicity and equi-percentile assumptions.

\subsubsection{Utilization of Covariates}
\label{sec:covs}

One approach to identify $P(T(0)>t|B(1)=1)$, is to utilize baseline
covariates $Z$, so that conditional on knowing $Z$, $B(1)$ provides no
further information on $T(0)$ (and vice versa). More formally the
requirement is that $T(0)$ and $B(1)$ are independent conditional on
covariates $Z$.  In formulas
\begin{equation}
  \label{eq:condind}
T(0) \perp B(1)  \; | \; Z.
\end{equation}

Using this assumption we have,
\begin{eqnarray*}
P(T(0)>t|B(1)=1)&=&\int P(T(0)>t|B(1)=1,Z)p(Z|B(1)=1)dZ\\ 
&=&\int P(T(0)>t|Z)p(Z|B(1)=1)dZ \\
&=&\int P(T(0)>t|X=0,Z)p(Z|B(1)=1)dZ \\
&=&\int P(T>t|X=0,Z)p(Z|X=1,B=1)dZ \\ \label{eq:ppr_form}
\end{eqnarray*}

The first equation follows by the law of total probability.  The
second equality follows by the assumption in \eqref{eq:condind}.  The
third equation follows by randomization \eqref{eq:rand} and the last
equation from the consistency assumption. This equation can be
estimated from the data by estimating $P(T>t|X=0,Z)$, and then
averaging it over the observed distribution of $Z$ for the population
of patients with $X=1$ and $B=1$ (\textit{i.e.} estimating
$p(Z|X=1,B=1)$ by its empirical distribution). We will discuss
specific methods for estimation in Section \ref{sec:ppr}. This
approach will be called ``predicted placebo response'' (PPR) in the
following.

Using the same arguments it follows that the probability density of
the event times in the subgroup is
$p(T(0)|B(1)=1)=\int p(T|X=0,Z)p(Z|X=1,B=1)dZ$. Furthermore,
{\small
\begin{eqnarray}
\int p(T|X=0,Z)p(Z|X=1,B=1)dZ &=& \int p(T,Z|X=0)\frac{p(Z|X=1,B=1)}{p(Z|X=0)}dZ \nonumber\\
&=& \int p(T,Z|X=0)\frac{p(B=1|Z,X=1)p(Z|X=1)p(X=1)}{p(Z|X=0)p(B=1,X=1)}dZ \nonumber\\
&\propto& \int p(T,Z|X=0)p(B=1|Z,X=1)dZ \label{eq:weight}
\end{eqnarray}}

The first two equations follow from the definition of conditional
probability. The last equation follows from randomization (because
$p(Z|X=0)=p(Z|X=1)$) and omitting all other terms that do not involve
$Z$. The result in (\ref{eq:weight}) is useful, because the observed
data on $T,Z$ with $X=0$ can be used to estimate $p(T(0)|B(1)=1)$ (and
$P(T(0)>t|B(1)=1)$) by utilizing the weights
\begin{equation}
  \label{eq:weights}
  w(Z)\propto p(B=1|X=1,Z).
\end{equation}
The weights $w(Z)$ can be estimated based on the patients on
treatment, using, for example, a logistic regression, or other
classification approaches. We will discuss specific methods for
estimation in Section \ref{sec:wpp}.  This approach will be called
``weighted placebo patients'' (WPP) in what follows.

\subsubsection{Utilizing Monotonicity and Equi-Percentile Assumptions}
\label{sec:bounds}

\begin{table}[h]
  \centering
  \begin{tabular}{c|c|c|c}
               & $B(0) = 1$        & $B(0) = 0$ \\ \hline 
    $B(1) = 1$ & $p_{11}$ & $p_{10}$ & $p_{1.}$ \\ 
    $B(1) = 0$ & $p_{01}$ & $p_{00}$ & $p_{0.}$ \\  \hline 
               & $p_{.1}$ & $p_{.0}$ &  
  \end{tabular}
  \caption{Principal strata}
  \label{tab:pi}
\end{table}

Table \ref{tab:pi} illustrates how the overall trial population can be
stratified into four subgroups according to their potential biomarker
outcomes under treatment and placebo. Here $p_{00}$, $p_{10}$,
$p_{01}$ and $p_{11}$ denote the probabilities to fall into the
relevant principal strata, $\{B(0)=i \wedge B(1)=j\}$ with $i,j \in
\{0,1\}$.  Each patient falls in exactly one subgroup so that
$p_{00}+p_{10}+p_{01}+p_{11}=1$. Note that this classification is not
known, because we only observe one of the two potential biomarker
measurements for every patient in the trial. Further let
$p_{.j}=p_{0j}+ p_{1j}$ and $p_{j.}=p_{j0}+ p_{j1}$ with a
$j\in\{0,1\}$ be the corresponding marginal probabilities of the
table, subject to $p_{0.}+p_{1.}=1$ and $p_{.0}+p_{.1}=1$. Note that
$p_{.j}$ and $p_{j.}$ can be estimated from the observed data on
placebo and active treatment, while the other probabilities are not
identifiable. To proceed further, a plausible assumption (depending on
the mechanistic understanding of the drug) is to assume that
\begin{equation}
  \label{eq:empty}
 p_{01}=0,
\end{equation}
\textit{i.e.}, there are no patients that would be biomarker
responders under placebo but not under treatment. This so-called
monotonicity assumption allows to identify that $p_{11}=p_{.1}$,
$p_{00}=p_{0.}$ and $p_{10}=p_{1.}-p_{.1}$. This is useful, because
(\ref{eq:pbob1}) can be expressed as

\begin{eqnarray}
  \label{eq:est2}
  P(T(0)>t|B(1)=1) &=& P(T(0)>t|B(1)=1) \nonumber \\
&=&\pi P(T(0)>t|B(1)=1 ,  B(0)=1)+ \nonumber \\
&&(1-\pi)P(T(0)>t|B(1)=1 ,  B(0)=0),
\end{eqnarray}

where $\pi=\frac{p_{11}}{p_{1.}}$.  Based on the assumption
(\ref{eq:empty}), the terms $\pi$ and
$P(T(0)>t|B(1)=1 ,  B(0)=1)=P(T>t|X=0,B=1)$ in (\ref{eq:est2}) are
identified from the trial data.

The term $P(T(0)>t|B(1)=1 , B(0)=0)$ however remains unidentified in a
parallel groups trial: Among the biomarker non-responders on placebo
it is not clear which would have responded under treatment.

Due to the monotonicity assumption, we can identify all proportions
$p_{ij}$ in Table \ref{tab:pi} and hence also the proportion of
patients $\tilde{\pi}$ among the placebo biomarker non-responders that
would have responded on treatment:
$\tilde{\pi}=\frac{p_{10}}{p_{.0}}=\frac{p_{1.}-p_{.1}}{p_{.0}}$.

The task is hence to identify the patients that would be biomarker
responders on treatment among the group of placebo biomarker
non-responders in order to estimate $P(T(0)>t|B(1)=1 , B(0)=0)$.  One
approach is to assume that the biomarker outcome $\beta(0)$ on placebo
contains information on the biomarker outcome under treatment
($\beta(1)$ and more specifically the event $B(1)=1$). Unfortunately
we do not have data relating $\beta(0)$ and $B(1)$, as every patient
received either placebo or the treatment, hence this relationship
cannot be estimated from the observed data and will be derived using
assumptions.

One simple approach is to rank patients according to their observed
placebo biomarker outcome. Then one could just select the fraction
$\tilde{\pi}$ of patients with the lowest value observed for
$\beta(0)$, and identify those as the ones that would be biomarker
responders on treatment. This type of assumption has in other contexts
been called equi-percentile equating (see \cite{rubi:1991}). This
approach can be criticized, because the ordering of potential
biomarker outcomes could be different under treatment and placebo. The
factors leading to a low biomarker value on placebo might be different
to those on treatment. Selecting the patients with low biomarker
outcome among the placebo biomarker non-responders (and identify those
as part of the stratum $B(1)=1 \wedge B(0)=0$) will hence tend to
overestimate the survival time under placebo and thus underestimate
the treatment effect.

A different idea is to include all patients among the placebo
biomarker non-responders to estimate $P(T(0)>t|B(1)=1 , B(0)=0)$.
This will also include patients with worse health-state, so one would
expect that the event time under placebo would get underestimated and
thus the treatment effect overestimated.

\begin{figure}[h!]
  \begin{center}
    \includegraphics[width=0.6\textwidth]{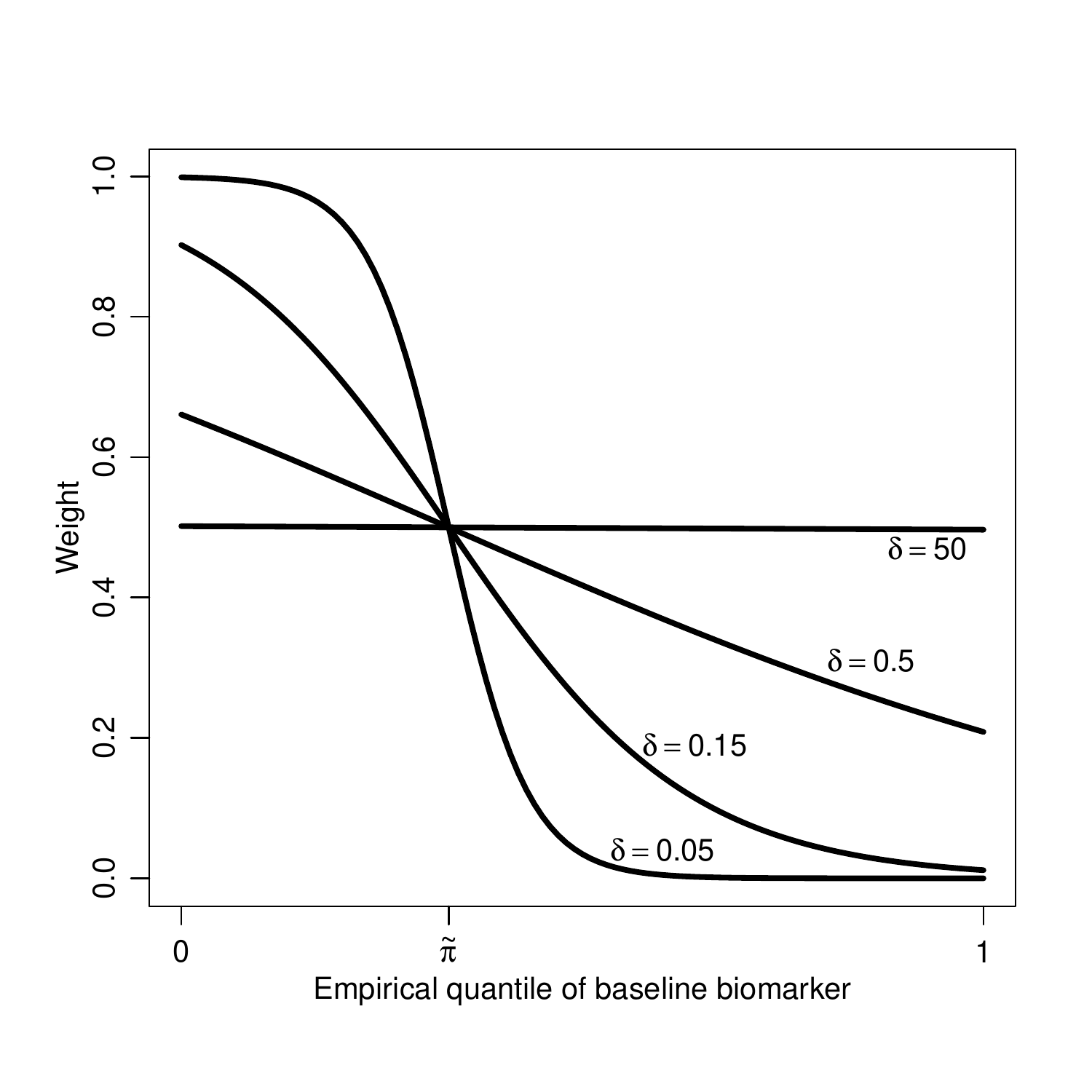}
  \end{center}
  \caption{Logistic weighting function with center $\tilde{\pi}$ and
    different $\delta$.}
  \label{fig:logistic}
\end{figure}

We propose an analysis that interpolates between these two extremes,
using a parameter, where patients with a lower rank and thus lower
observed $\beta(0)$ get a higher weight than patients with higher
$\beta(0)$. We propose to use the weight function
$\omega_i=1-\frac{1}{1-\exp(-(\tau_i-\tilde{\pi}))/\delta)}$ for
patient $i$, where $\tau_i$ is the empirical quantile for $\beta(0)$
of patient $i$, \textit{i.e.}, if there are $\eta$ observations among
the placebo biomarker non-responders, the patient with the $i$-th
ordered observation of $\beta(0)$ will have an empirical quantiles of
$\tau_i=i/(\eta+1)$. See Figure \ref{fig:logistic} for an illustration
of this weight function for different $\delta$ values. If
$\delta \rightarrow 0$ this approach is equivalent to using the
equi-percentile equating assumption (the $\tilde{\pi}$ patients with
lowest $\beta(0)$ will receive a weight of 1, while all others receive
a weight of 0). Letting $\delta \rightarrow \infty$ corresponds to
weighting patients equally.

To summarize the assumptions underlying this approach, we use the
monotonicity assumption (\ref{eq:empty}) and the relaxed
equi-percentile assumption, based on a logistic weighting function.
We introduced a parameter $\delta$ that interpolates between two
extremes: For $\delta \approx 0$ this analysis might tend to
underestimate the treatment effect while for
$\delta \rightarrow \infty$ this analysis will often tend to
overestimate the treatment effect. This specific approach will be
called analysis based on monotonicity and equi-percentile assumption
(MEA) in what follows.
\section{Estimation}
\label{sec:est}

Different statistical models can be utilized to estimate the
quantities (\ref{eq:estimand1}) and (\ref{eq:estimand2}). Bayesian
methods have the appeal of directly providing an uncertainty
assessment, for non-Bayesian methods bootstrap approaches can be
utilized to perform inference. We focus here on semiparametric methods
and utilize bootstrapping for inference. Nevertheless other approaches
could equally be used for estimation.

Estimation of $P(T(1) > t|B(1) = 1)$ is straightforward by estimating
$P(T>t|X=1, B=1)$ from the observed data, for example, using the
Nelson-Aalen estimator.

In Section \ref{sec:methods} three methods were described to enable
estimation of $P(T(0) > t|B(1) = 1)$: The PPR and WPR approaches, both
utilizing baseline covariates and the MEA utilizing the monotonicity
and equi-percentile assumptions, in the following sections, we
summarize the statistical analyses that can be performed to estimate
the quantities needed for each approach.

\subsection{Predict placebo response (PPR)}
\label{sec:ppr}

The PPR approach requires estimation of $P(T > t|X=0, Z)$ (see
\eqref{eq:ppr_form}), \textit{i.e.} estimation of the survival in the
placebo arm based on covariates $Z$. This can be done by fitting a Cox
regression in the placebo arm. An estimate for $p(Z|X=1,B=1)$ can be
derived by using the empirical distribution of covariates for the
biomarker responders on the treatment arm. To derive an estimate for
\eqref{eq:pbob1} the observed covariates $Z$ for biomarker responders
on treatment will hence be used to predict a survival curve for every
patient. This can be done using the Breslow estimator of the baseline
survival function. The estimate of (\ref{eq:pbob1}) is then the
average of these predicted survival curves. The main modelling
assumption in this approach is that covariates enter multiplicatively
on the hazard rate according to the Cox proportional hazards model.

\subsection{Weight placebo patients (WPP)}
\label{sec:wpp}

The WPP approach requires a model to determine the weights $w(Z)$ from
(\ref{eq:weights}). For that purpose a logistic regression on the
treatment arm is fitted, where the outcome is the binary indicator of
the event $B=1$ (being a biomarker responder) and the covariates are
$Z$. Then using this model, for each patient on the placebo arm the
probability for $B=1$ is predicted. These probabilities are used as
weights in the weighted estimation of the survival function using a
weighted Nelson-Aalen estimator in the placebo group, which is thus
the estimate of $P(T(0) > t|B(1) = 1)$.  The main modelling assumption
here originates from the model assumed for determination of $w(Z)$,
\textit{i.e.} the logistic regression.

\subsection{Monotonicity and Equi-Percentile Assumption (MEA)}
\label{sec:ps}

The MEA approach requires estimation of $\pi$ and $\tilde{\pi}$ (see
Section \ref{sec:bounds}). This will be done directly from the
observed proportions based on the formulas in Section \ref{sec:bounds}
that follow from the monotonicity assumption (\ref{eq:empty}). For the
quantity $P(T(0)>t|B(1)=1,B(0)=1)$ we have from the monotonicity
assumption that
$P(T(0)>t|B(1)=1,B(0)=1)=P(T(0)>t|B(0)=1)=P(T>t|X=0,B=1)$ where the
last equation follows by consistency. For estimation of this quantity
the Nelson-Aalen estimator will be used.  The quantity
$P(T(0)>t|B(1)=1,B(0)=0)$ will be estimated in the group of biomarker
non-responders on placebo, weighted according the approach outlined in
Section \ref{sec:bounds}, where the weights are derived based on the
biomarker value on placebo and the logistic function with center
$\tilde{\pi}$ and sensitivity parameter $\delta$ that is chosen
independent of the observed data.  A weighted Nelson-Aalen estimator
will be used.

\section{Simulations}
\label{sec:sims}

The purpose of this section is to evaluate the theory developed in
Section \ref{sec:methods} for a particular data generating true
model.

We will generate data for a parallel groups, event-driven randomized
trial to compare active treatment against placebo. The simulation is
loosely motivated by the CANTOS trial mentioned in Section
\ref{sec:intro}, even though a few details used here are different. It
is assumed that around 20\% of the patients will have an event by year
5. The analysis will be performed, once 850 events have been observed
in total. Patients that did not have an event by this calendar time
will be censored. The number of 850 events is chosen as this is
approximately the number needed to detect a log-hazard ratio of 0.8
based on the log-rank test with significance level $0.05$ and power
$0.9$.
Recruitment will be simulated according to a homogeneous Poisson
process, that is, enrollment is assumed to be linear increasing over
time (uniformly distributed entry times). The yearly recruitment rate
is set to be 1500 patients.  For every patient we utilize two baseline
covariates, $Z_0$ and $Z_1$ (in a real situation $Z$ would of course
be higher-dimensional). Here, $Z_0$ is assumed to be the biomarker
value at baseline and $Z_1$ a general covariate that has a strong
prognostic effect on the event time, and small effect on the
post-baseline biomarker. For $Z_0$ and $Z_1$ we assume a bivariate
normal distribution with means 0, standard deviations 1 and
correlation of 0.25. The post-baseline biomarker level $\beta$ for
patient $i$ is then simulated from the following linear model
\begin{equation}
  \label{eq:beta-mod}
 \beta_i\sim N(\mu_i,1)\;\mathrm{with}\;\mu_i=\alpha_0+\alpha_1X_i +\alpha_2Z_{0,i}+\alpha_3Z_{1,i}
\end{equation}
where $Z_{0,i},Z_{1,i}$ and $X_i$ are the observed baseline covariates
and the treatment indicator for patient $i$. In the simulations the
parameters $\boldsymbol{\alpha}=(\alpha_0,\alpha_1,\alpha_2,\alpha_3)$
will be chosen to represent a typical situation where there is a
strong effect of treatment and baseline biomarker value $Z_0$ on the
post-baseline biomarker, but only a small effect of the covariate
$Z_1$. The specific values assumed are $\alpha_0=1$, $\alpha_1=-1.75$,
$\alpha_2=0.5$ and $\alpha_3=0.1$.

The event time $T_i$ for patient $i$ will be generated based on the
post-baseline biomarker value $\beta_i$ for every
patient as
\begin{equation}
  \label{eq:expo}
  T_i \sim \mathrm{Exponential}(\lambda_i)\;\mathrm{with}\;
  \lambda_i= \exp(\gamma_0 + \gamma_1Z_{0,i} + \gamma_2Z_{1,i} + \gamma_3X_i + \gamma_4\beta_i +
  \gamma_5\beta_i X_i).
\end{equation}

The parameters
$\boldsymbol{\gamma}=(\gamma_0,\gamma_1,\gamma_2,\gamma_3,\gamma_4,\gamma_5)$
are chosen as follows: $\gamma_0$ is chosen so that at 5 years an
event rate of 20\% is achieved (if all other covariates would be
0). The parameters $\gamma_1$ and $\gamma_2$ represent the prognostic
effect of the baseline variables, for $\gamma_1$ a value of
$-\log(0.95)$ will be chosen, indicating a small effect the baseline
biomarker level and for $\gamma_2$ $-\log(0.5)$ indicating a strong
effect of this prognostic covariate.

The parameters $\gamma_3,\gamma_4$ and $\gamma_5$ describe the
treatment effect: $\gamma_3$ denotes the effect of treatment
independent of the post-baseline biomarker level, $\gamma_4$ is the
effect of the post-baseline biomarker level $\beta$, while $\gamma_5$
determines, how much the treatment effect is modified by the
post-baseline biomarker level. Different scenarios will be evaluated
for $\gamma_3,\gamma_4$ and $\gamma_5$, corresponding to different
assumptions on how the treatment effect is generated: (i) for
$\gamma_3\neq0$, $\gamma_4=\gamma_5=0$ the post-baseline biomarker
value has no effect on the outcome, and the treatment works only by
other mechanisms, (ii) for $\gamma_4\neq 0$, $\gamma_3=\gamma_5=0$ all
the treatment effect is achieved through the modification of the
biomarker, and (iii) $\gamma_3=0$ and $\gamma_4,\gamma_5\neq 0$
corresponds to situation (ii), but the treatment effect is modified by
the post-baseline biomarker level. In each case the parameters
$\gamma_3,\gamma_4,\gamma_5$ are chosen such that difference in the
average log hazard rates between treatment and placebo group is equal
to $\log(0.8)$, to ensure that the overall ``treatment effect'' is
similar across the scenarios. The exact parameters values are given in
Table \ref{tab:gamma} in Appendix \ref{sec:true-data-generating}. A
detailed algorithm of how the event-driven trials are generated, is
given in Appendix \ref{sec:data-gener-event}.

For the simulation we will set the biomarker threshold at $0$,
\textit{i.e.}, we focus on the subgroup of patients that achieve a
post-baseline biomarker value less than $0$ on treatment. From model
(\ref{eq:beta-mod}) for the post-baseline biomarker and with
parameters as in Appendix \ref{sec:true-data-generating}, the
probability to achieve $\beta < 0$ on treatment is around $75\%$ and
around $19\%$ on placebo.

We estimate the difference in restricted survival time
$\Delta_{RMST,t^*}$ with $t^*=5$.  For estimation of the difference in
survival $\Delta(t)$ we utilize $t=2,5$. 

We compare six approaches: PPR, WPP and MEA (with $\delta=0.05$ and
$\delta=50$) and two further analyses which are simple and seemingly
intuitive to do, but do not target the estimand of interest. The first
additional method estimates the placebo survival curve in the subgroup
of interest by utilizing the complete placebo group (called
NAIVE\textunderscore FULLPBO in what follows). The second approach
only utilizes placebo patients, which have $\beta < 0$ on placebo
(NAIVE\textunderscore THRES). As discussed in the introduction, the
NAIVE\textunderscore FULLPBO approach is expected to over-estimate the
treatment effect (as the baseline population on the combined placebo
group is ``less healthy''), while for the NAIVE\textunderscore THRES
approach one would expect that it underestimates the treatment effect
(as the baseline population of patients on the placebo arm that reach
the threshold is ``healthier'').

To evaluate how well the different approaches estimate the estimands
of interest (difference in survival curve and restricted mean survival
time), we need to derive the ``true'' survival curves on treatment and
placebo in the subgroup of interest. In Appendix
\ref{sec:calc-true-surv} this is explained in detail.

\begin{figure}[h!]
  \begin{center}
    \includegraphics[width=0.95\textwidth]{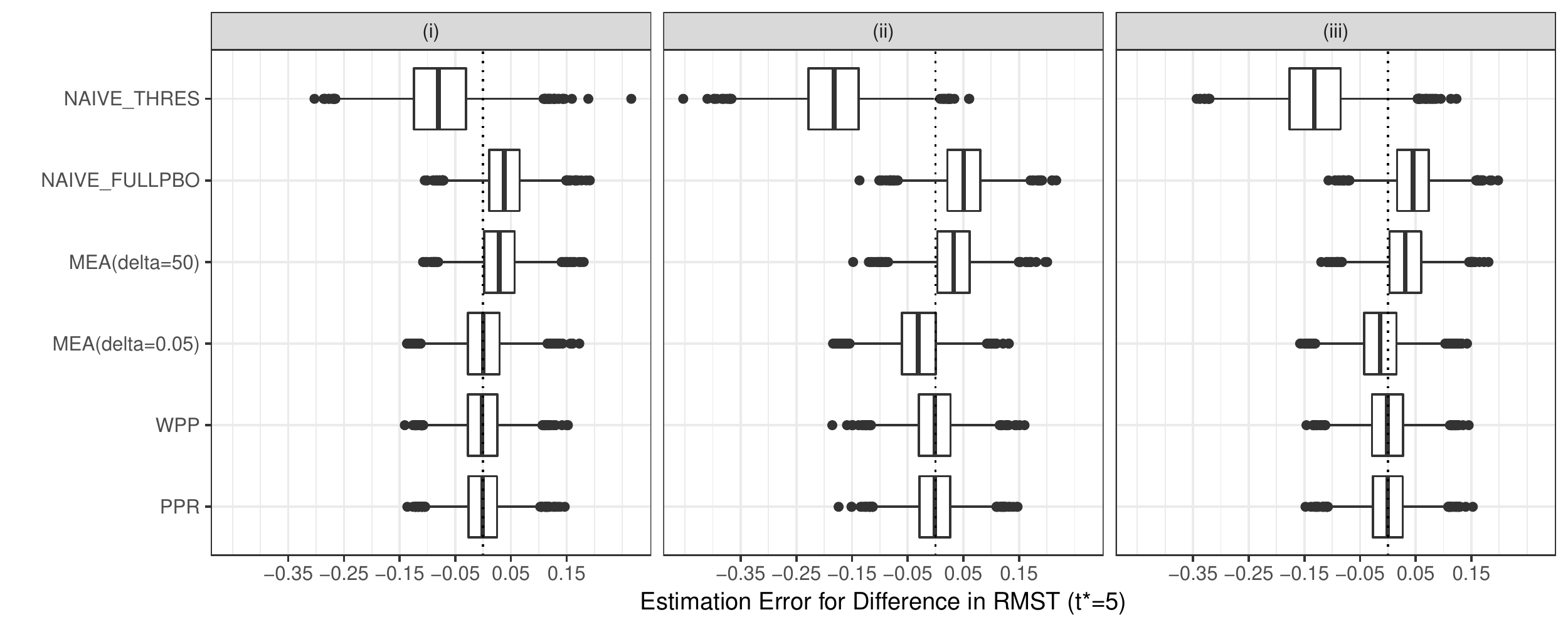}
  \end{center}
  \caption{Boxplot of difference between true and estimated difference
    in the mean restricted survival time (RMST) up to year 5.}
  \label{fig:rmst}
\end{figure}

In each scenario 5000 simulations were performed and Figure
\ref{fig:rmst} displays boxplots for the estimation error of the
difference in the restricted mean survival time up to year 5
($\Delta_{RMST,5}$). As expected the naive approaches systematically
over- (NAIVE\textunderscore FULLPBO) or underestimate
(NAIVE\textunderscore THRES) the difference in restricted mean
survival times, with the NAIVE\textunderscore THRES approach having a
much larger variability (due to the small sample size on placebo).

For the MEA approach one can see that for our simulation settings the
method with $\delta=50$ one obtains estimates that tend to
overestimate the treatment effect, as expected. The performance for
$\delta=0.05$ is adequate in scenario (i), where the outcome is
independent of the post-baseline biomarker. For scenarios (ii) and
(iii) the treatment effect is underestimated on average. This is
because the equi-percentile assumption is violated in these scenarios
(see also Appendix \ref{sec:calc-true-surv} on how the true survival
differences were calculated). The WPP and PPR perform well across all
scenarios. This is expected as these approaches utilize the
information on the true covariates (and in this sense also the true
simulation model).

Overall similar results can also be observed for estimation of the
survival differences $\Delta(t=2)$ and $\Delta(t=5)$, see Figures
\ref{fig:sd2} and \ref{fig:sd5} in the Appendix.

Note that for the considered simulation setting is simple, but the
purpose here was to investigate our semiparametric procedures in this
case, where naive approaches already fail.

\section{Data application}
\label{sec:appl}

For illustration we analyse a data set simulated under scenario (iii)
above. In the simulated data set 850 events were reached after 5.9
years. The mean follow-up time for patients was around 3.9 years at
that time.

\begin{figure}[h!]
  \begin{center}
    \includegraphics[width=0.7\textwidth]{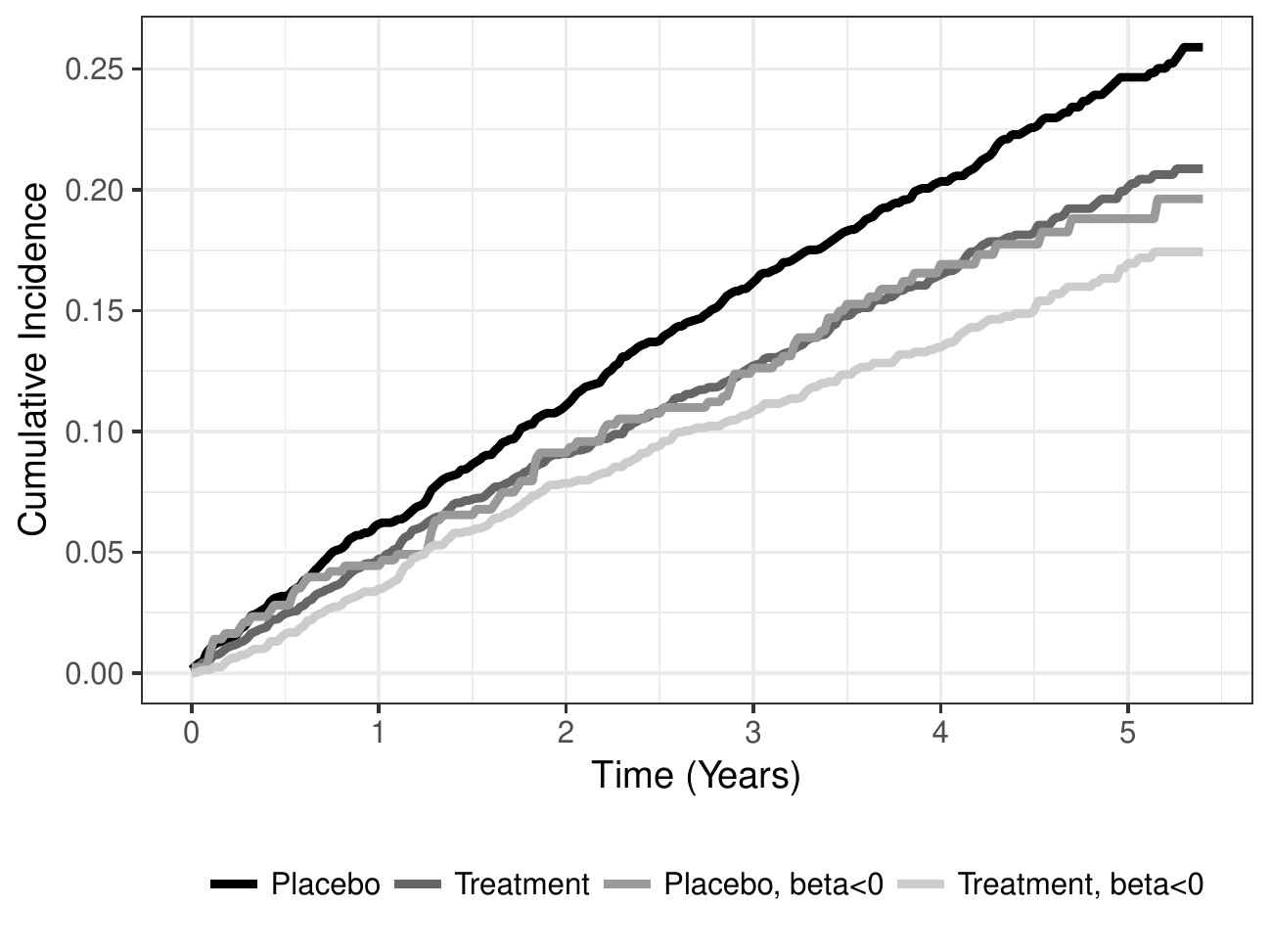}
  \end{center}
  \caption{Observed cumulative incidence rates.}
  \label{fig:cum-incid}
\end{figure}

Figure \ref{fig:cum-incid} shows four observed cumulative incidence
rates 1$-$(estimated survival function), estimated using the
Nelson-Aalen estimator. Two curves correspond to the overall estimates
in the placebo and treatment groups. The other two curves correspond
to patients on placebo and treatment with $\beta < 0$.  About $20\%$
of the patients on placebo reached this value and about $76\%$ of the
patients on treatment.  Figure \ref{fig:cum-incid} shows that there
were less events on the active treatment and also that a low biomarker
post-baseline leads to a smaller event rate.

\begin{figure}[h!]
  \begin{center}
    \includegraphics[width=0.95\textwidth]{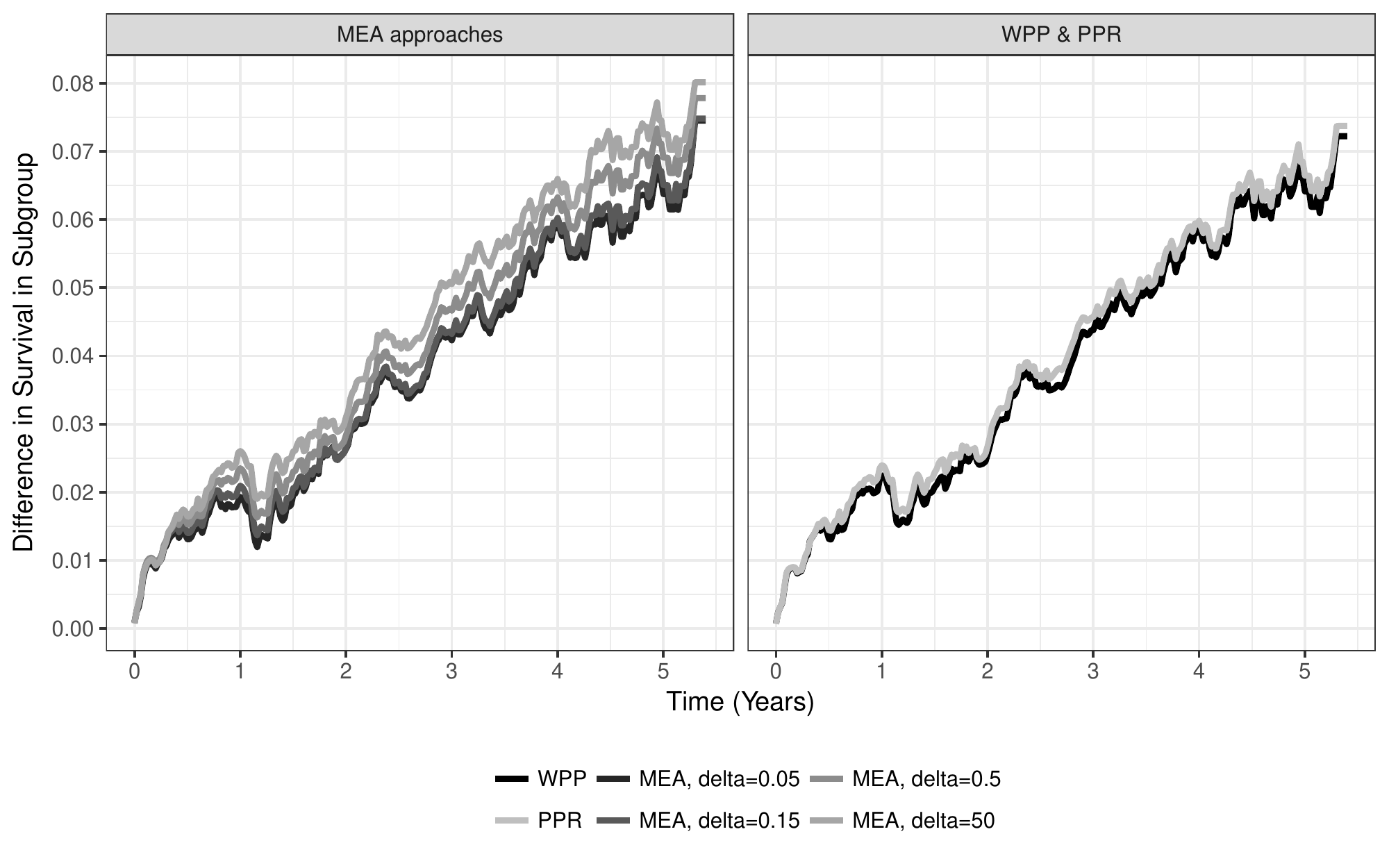}
  \end{center}
  \caption{Estimated difference between the survival curves.}
  \label{fig:surv-diff}
\end{figure}

Figure \ref{fig:surv-diff} shows six approaches to estimate the
difference in the cumulative incidence rate. The first two are the PPR
and WPP approaches discussed in Section \ref{sec:methods}. In addition
we show four principal stratification approaches with the different
$\delta$ values shown in Figure \ref{fig:logistic}. Note that all six
approaches only differ in the way that the placebo survival curve is
estimated. The estimation of the survival curve in the subgroup of
interest under treatment is the same for all approaches. The results
are quite consistent with the simulation results in the previous
section in the sense that the PPR and WPP approaches lead to quite
similar results and the MEA approaches estimate an increased treatment
effect with increasing $\delta$ as expected.

Table \ref{tab:bres} presents numerical results for the PPR, WPP and
MEA approaches for estimation of the survival differences $\Delta(2)$,
$\Delta(5)$ and $\Delta_{RMST,5}$. Confidence intervals have been
calculated using sampling with replacement stratified by treatment
group (non-parametric bootstrap).

\begin{table}
\center
  \begin{tabular}{ |l|c|c|c| } 
    \hline
Method  &  $\Delta(2)$ &  $\Delta(5)$ & $\Delta_{RMST,5}$ \\ \hline
WPP & 0.026 (0.01,0.04) & 0.064 (0.039,0.09) & 0.183 (0.114,0.248)\\
PPR & 0.027 (0.012,0.042) & 0.066 (0.043,0.093) & 0.191 (0.123,0.256)\\
MEA, $\delta=0.05$ & 0.026 (0.01,0.042) & 0.064 (0.037,0.09) & 0.178 (0.103,0.245)\\
MEA, $\delta=0.15$ & 0.027 (0.011,0.042) & 0.065 (0.039,0.09) & 0.182 (0.111,0.249)\\
MEA, $\delta=0.5$ & 0.028 (0.013,0.044) & 0.069 (0.045,0.093) & 0.196 (0.128,0.265)\\
MEA, $\delta=50$ & 0.031 (0.016,0.046) & 0.073 (0.048,0.096) & 0.211 (0.145,0.28)\\\hline
  \end{tabular}
\caption{Point estimates with 90\% bootstrap quantile confidence intervals for the
  quantities of interest.}
\label{tab:bres}
\end{table}

One can see that in all cases and for all estimands of interest the
confidence interval excludes 0, so a treatment effect is concluded in
this setting, which is the right decision based on the simulation
scenario upon which the data were generated.

\section{Conclusions}
\label{sec:concl}

In this paper we considered estimation of the treatment effect in a
subgroup defined by reaching a post-baseline biomarker measurement on
treatment. This is challenging, because in a parallel groups trial we
do not observe the event time that these patients would have had, had
they been randomized to the placebo group. Three approaches are
proposed based on different causal identifying assumptions. Two
approaches (PPR and WPP) are based on a conditional independence
assumption utilizing baseline covariates, stating that given a set of
baseline covariates the potential biomarker outcome under treatment
and the potential event time under placebo are independent. Utilizing
this, one can use placebo information to estimate the placebo response
for the patients in the subgroup of interest. The third approach (MEA)
is based on principal stratification and utilizes the monotonicity and
a relaxed equi-percentile assumption with a logistic weighting
function based on a sensitivity parameter $\delta$. The approaches
have been evaluated in a simulation study, where the proposed
approaches show good performance compared to more naive
approaches. Finally the methodology has been illustrated on a single
simulated data set.

In this paper we estimated the survival difference at a specified time
and on estimation of the mean restricted survival time, both
quantities are rather easy to interpret. In many clinical applications
the hazard ratio is the measure to compare time-to-event endpoints,
with the underlying assumption that the hazards are proportional. In
our setting we do not make the assumption of proportional hazards and
estimate the survival curves separately for placebo and treatment.
Hence it is not immediately obvious how to summarize the ratio of
hazard functions into a single hazard ratio. A graphical approach
would be to plot the ratio of the cumulative hazard functions, which
is easily obtainable from the already determined survival functions,
to see whether values fluctuate around a particular value. In cases
like the simulated example above (see Section \ref{sec:appl}), where
there is a rather low event rate, and the cumulative hazard functions
are approximately linear (like in Figure \ref{fig:cum-incid}), one
approach is to fit a simple exponential ($\exp(-\lambda t)$) to the
survival curves and then derive an approximate hazard ratio based on
the ratio of the fitted rates.

\hspace{0.5cm}\\
\hspace{-0.6cm}\textbf{Acknowledgments}\\
The authors would like to thank Daniel Scharfstein for sharing ideas
and discussions on the topic, as well as feedback on early versions of
the methods described in this paper.  We would also like to thank
Heinz Schmidli, Simon Wandel and Nathalie Ezzet, who provided comments
on the methods and earlier versions of the manuscript.

\bibliographystyle{agsm}

\newpage
\begin{appendix}

\section{Simulations}
\label{sec:sim-details}

\subsection{True data-generating parameters}
\label{sec:true-data-generating}

For the parameter $\boldsymbol{\gamma}$ the scenarios outlined in
Table \ref{tab:gamma} will be utilized.

\begin{table}
\center
  \begin{tabular}{ |l|c| } 
    \hline
    Scenario &  Values \\ \hline
    (i) $\gamma_3\neq0$, $\gamma_4=\gamma_5=0$ & $\gamma_3=\log(0.8)$  \\ 
    (ii) $\gamma_3=\gamma_5=0$, $\gamma_4\neq 0$  & $\gamma_4=0.1275$  \\
    (iii) $\gamma_3=0$, $\gamma_4,\gamma_5\neq 0$ & $\gamma_4=0.06375, \gamma_3=0.1489$  \\ \hline
  \end{tabular}
\caption{Parameter values for $\gamma$ utilized in the three
  simulation scenarios.}
\label{tab:gamma}
\end{table}

\subsection{Data generation for event-driven trial}
\label{sec:data-gener-event}

The following steps explain how the survival data were generated in
the simulation study.

\begin{itemize}
\item Input\\
  $\lambda_0$: enrollment rate per year\\
  $p_{\mathcal{Y}}$ and $\mathcal{Y}$: A fraction of $p_{\mathcal{Y}}$
  of the subjects on placebo will have an event
  by year ${\mathcal{Y}}$ \\
  $n_e$ number of events\\
  $\boldsymbol{\alpha},\boldsymbol{\gamma}$: parameters for the models
  of the post-baseline biomarker and the event-time
\item Calculate $N^*=[n_e/p_{\mathcal{Y}}]$, the approximate number needed to
recruit to achieve $n_e$ events by year ${\mathcal{Y}}$.
\item For each patient simulate the following random variables
  (\textit{i.e.} $N^*$ in total)
  \begin{itemize}
  \item Generate $N^*$ i.i.d. exponential random variates with hazard
    rate $\lambda_0$ and sort them in increasing order, to obtain the
    recruitment times for every patient
  \item Generate baseline variables $(Z_0,Z_1)\overset{i.i.d.}{\sim} MVN(\mu, \Sigma)$
    with mean vector equal to 0, marginal variances 1 and correlation
    0.25.
  \item treatment indicator $X\overset{i.i.d.}{\sim} B(1,0.5)$
  \item post-baseline biomarker value $\beta$ according to the linear
    model (\ref{eq:beta-mod}) with parameters $\boldsymbol{\alpha}$
    and covariates values as generated in the previous steps.
  \item the event time from the exponential model (\ref{eq:expo}) with
    parameters $\boldsymbol{\gamma}$ and covariates values as
    generated in the previous steps.
  \end{itemize}
  \item Calculate the calendar times for the event times
    (\textit{i.e.} for every patient add recruitment and event time).
  \item Determine by which calendar time 850 patients had the
    event. Remove all patients that were recruited after this calendar
    time. The event times of patients that had their event-time after
    this calendar time are right-censored with this as their censoring
    time. For analysis translate all times back to the study time.
\end{itemize}

\subsection{Calculation of true survival curves in subgroup of
  interest}
\label{sec:calc-true-surv}

In the data-generating model used in our simulations (Equations
(\ref{eq:beta-mod}), (\ref{eq:expo})) the subgroup of patients of
interest (\textit{i.e.}  those that achieve $\beta < 0$ on treatment)
can explicitly be characterized in terms of the baseline covariates
$Z_0$ and $Z_1$: The joint distribution of $Z_0$ and $Z_1$ conditional
on $\beta < 0$ and $X=1$ can be simulated by simulating the joint
distribution given $X=1$, and then removing the observations where
$\beta \geq 0$.

Having obtained this joint distribution for $Z_0$, $Z_1$ and $\beta$
in the subgroup of interest (patients with $\beta < 0$), these values
can be plugged in the true exponential survival curve (see
(\ref{eq:expo})) and the resulting survival curves averaged, to obtain
the population survival curve under treatment. To obtain the
population survival curve under placebo a biomarker outcome under
placebo is first simulated for each patient based on their covariates
$Z_0$, $Z_1$ (according to (\ref{eq:beta-mod})) and then $Z_0$, $Z_1$
and the simulated biomarker value under placebo are plugged in the
true exponential survival curve. The resulting survival curves are
then averaged, to obtain the population survival curve under
treatment.

\subsection{Additional Simulation Results}
\label{sec:add-results}

\begin{figure}[h!]
  \begin{center}
    \includegraphics[width=0.95\textwidth]{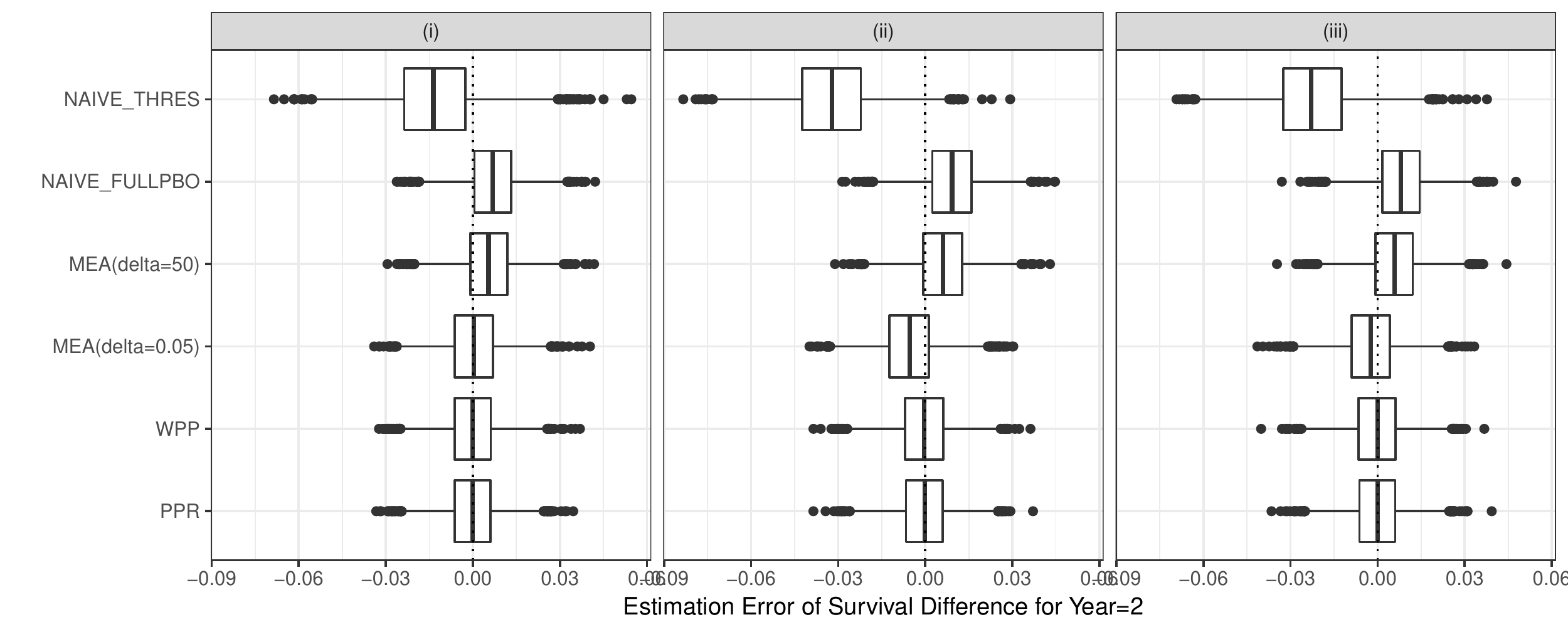}
  \end{center}
  \caption{Estimation error for estimation of the difference in the
    survival curves at time point 2.}
  \label{fig:sd2}
\end{figure}

\begin{figure}[h!]
  \begin{center}
    \includegraphics[width=0.95\textwidth]{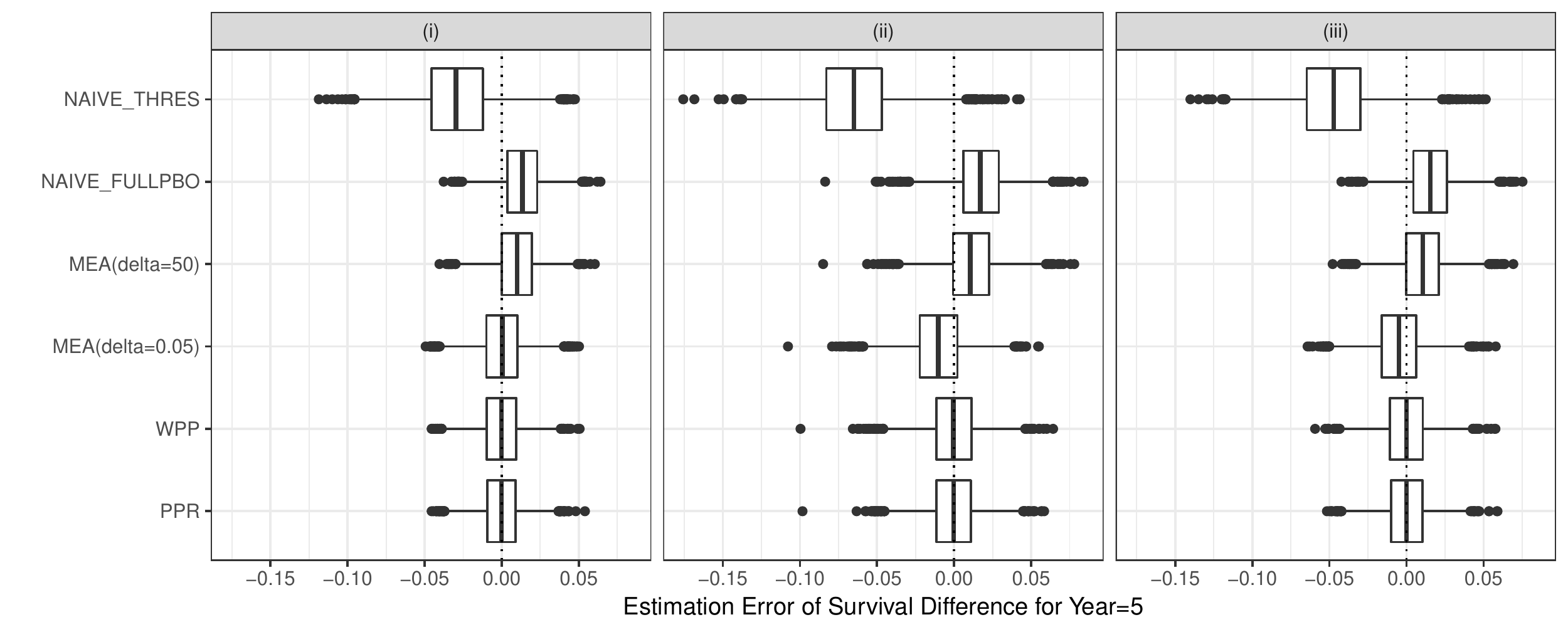}
  \end{center}
  \caption{Estimation error for estimation of the difference in the
    survival curves at time point 5.}
  \label{fig:sd5}
\end{figure}

\section{Events before $\mathcal{T}$ and missing post-baseline
  biomarker measurements}
\label{sec:ev}

The time-point $\mathcal{T}$ of measuring the post-baseline biomarker,
should be relatively short after initiation of treatment.  A subgroup
defined based on a biomarker measurement at a time-point far away from
baseline, will not be of clinical interest. Thus in general few events
will be expected before $\mathcal{T}$, and unlikely to influence the
overall analysis much. Nevertheless in many practical situations there
will be some events and in this section we will discuss how to
formally handle these situations. Another related issue that we will
discuss, are missing post-baseline biomarker values (\textit{i.e.}
situations where a biomarker value could have been measured at time
$\mathcal{T}$ but was not).

An important consideration in handling events before time
$\mathcal{T}$ in the case of composite events is to distinguish
between death and non-fatal events. Let the time to death be $T_D$. In
general the subgroup of interest should then be defined by
$B(1) = 1 \;\mathrm{and}\; T_D(1) > \mathcal{T}$, \textit{i.e.}, the
subgroup of patients, who would, on treatment, be alive and biomarker
responders.  An assumption that is plausible in some situations is to
assume that $T_D(1) > \mathcal{T}$ implies $T_D(0) > \mathcal{T}$ and
vice-versa. That is, patients not dying under treatment before time
$\mathcal{T}$ would also not have died under placebo (and vice versa).
This assumption would justify an analysis, where patients dying before
$\mathcal{T}$ are excluded from the analysis and approaches suggested
in Section \ref{sec:methods} could be performed as they are
described. If it is assumed that the population of patients dying
before time $\mathcal{T}$ is different between treatment and placebo,
the analyses would need to be modified, to identify/weight patients on
the placebo arm, according to whether they would have died on the
treatment arm until time $\mathcal{T}$.

In principle non-fatal events could be handled in exactly the same
way. One could define the subgroup by
$B(1) = 1 \;\mathrm{and}\; T(1) > \mathcal{T}$, \textit{i.e.}, the
subgroup of patients, who would, on treatment, be event-free and
biomarker responders. Whether or not this is more relevant than the
subgroup $B(1) = 1 \;\mathrm{and}\; T_D(1) > \mathcal{T}$ is a
clinical and pharmacological question: Are non-fatal events likely due
to the fact that the drug ``does not work'' (\textit{i.e.} supporting
the decision of stopping the treatment after the event, which would
mean it is more relevant to focus on the subgroup
$B(1) = 1 \;\mathrm{and}\; T(1) > \mathcal{T}$), or is it too early to
say that, because the drug effect did not fully materialize by time
$\mathcal{T}$. If it is considered relevant to continue treating
patients even in case of an event up to time $\mathcal{T}$ the more
relevant subgroup would be given by
$B(1) = 1 \;\mathrm{and}\; T_D(1) > \mathcal{T}$.  In this situation
the approach would be to include the events before $\mathcal{T}$ in
the analysis.

A different issue is that there will always be patients, where the
post-baseline biomarker measurement is missing, but in principle
measurable. It is not appropriate to remove these patients from the
analysis, as these might be systematically different from the
population of patients, where the biomarker measurement is available.
One way of approaching this, is to impute missing post-baseline
biomarker values, using for example the baseline biomarker level as
well as other covariates to impute the post-baseline biomarker value
on treatment.  Given each completed multiply-imputed data set, the
rest of the analysis would be conducted as described in Sections
\ref{sec:methods} and \ref{sec:est} and at the end appropriately
combined.

\end{appendix}

\end{document}